\pdfoutput=1

\documentclass[sigconf]{acmart}

\AtBeginDocument{%
  \providecommand\BibTeX{{%
    \normalfont B\kern-0.5em{\scshape i\kern-0.25em b}\kern-0.8em\TeX}}}

\copyrightyear{2020}
\acmYear{2020}
\setcopyright{acmlicensed}
\acmConference[FAT* '20]{Conference on Fairness, Accountability, and Transparency}{January 27--30, 2020}{Barcelona, Spain}
\acmBooktitle{Conference on Fairness, Accountability, and Transparency (FAT* '20), January 27--30, 2020, Barcelona, Spain}
\acmPrice{15.00}
\acmDOI{10.1145/3351095.3372877}
\acmISBN{978-1-4503-6936-7/20/02}

\begin{document}

\title{Awareness in Practice: Tensions in Access to Sensitive Attribute Data for Antidiscrimination}

\author{Miranda Bogen}
\affiliation{%
\institution{Upturn}}
\authornote{Author was affiliated with Upturn at time of writing.}
\email{mirandabogen@gmail.com}

\author{Aaron Rieke}
\affiliation{%
\institution{Upturn}}
\email{aaron@upturn.org}

\author{Shazeda Ahmed}
\affiliation{%
\institution{University of California, Berkeley}}
\authornote{Author was a Fellow at Upturn at time of writing.}
\email{shazeda@ischool.berkeley.edu}

\renewcommand{\shortauthors}{Bogen, Rieke, and Ahmed}

\begin{abstract}
Organizations cannot address demographic disparities that they cannot see. Recent research on machine learning and fairness has emphasized that awareness of sensitive attributes, such as race and sex, is critical to the development of interventions. However, on the ground, the existence of these data cannot be taken for granted.

This paper uses the domains of employment, credit, and healthcare in the United States to surface conditions that have shaped the availability of sensitive attribute data. For each domain, we describe how and when private companies collect or infer sensitive attribute data for antidiscrimination purposes. An inconsistent story emerges: Some companies are required by law to collect sensitive attribute data, while others are prohibited from doing so. Still others, in the absence of legal mandates, have determined that collection and imputation of these data are appropriate to address disparities.

This story has important implications for fairness research and its future applications. If companies that mediate access to life opportunities are unable or hesitant to collect or infer sensitive attribute data, then proposed techniques to detect and mitigate bias in machine learning models might never be implemented outside the lab. We conclude that today's legal requirements and corporate practices, while highly inconsistent across domains, offer lessons for how to approach the collection and inference of sensitive data in appropriate circumstances. We urge stakeholders, including machine learning practitioners, to actively help chart a path forward that takes both policy goals and technical needs into account.
\end{abstract}


\maketitle

\section{Introduction}
Statistical models, including those created with machine learning, can reproduce biases in the historical data used to train them. As powerful institutions increase their reliance upon these models to automate decisions that affect people's rights and life opportunities, researchers have begun developing new techniques to help detect and address these biases. The real-world implementation of these techniques could be an essential part of ensuring the continued viability of civil and human rights protections.

Many machine learning fairness practitioners rely on awareness of sensitive attributes---that is, access to labeled data about people's race, ethnicity, sex, or similar demographic characteristics---to test the efficacy of debiasing techniques or directly implement fairness interventions. A significant body of research presumes the modeler has ready access to data on these characteristics as they build and test their models \cite{Dwork2012, Feldman2015, Chen2019}. The need for this data is plain to see. As a 2003 analysis of racial disparities in healthcare powerfully concluded: "The presence of data on race and ethnicity does not, in and of itself, guarantee any subsequent actions ... to identify disparities or any actions to reduce or eliminate disparities that are found. The absence of data, however, essentially guarantees that none of those actions will occur." \cite{AHRQ2012}

Increasingly, companies that utilize machine learning are being asked to detect and address bias in their products. But they are not the first to grapple with these issues. This paper explores the legal and institutional norms surrounding the collection, inference, and use of sensitive attribute data in three key corporate domains. This analysis has significant implications for machine learning fairness research: If private institutions that mediate access to life opportunities are unable or hesitant to collect or infer sensitive attribute data, then emerging awareness-based techniques to detect and mitigate bias in machine learning models might never be implementable in real-world settings. 

Notably, this paper does not discuss complex and important questions about \emph{how} "fairness" should be measured or addressed, recognizing that definitions are manifold \cite{Narayanan18}. Rather, we make a simpler point: If sensitive attribute data are not available, interventions that rely on them will be severely impaired.

We conduct this exploration through the lens of U.S. civil rights law in the domains of credit, employment, and healthcare. For each domain, we describe when and how private companies collect or infer sensitive attribute data to pursue antidiscrimination goals. These are not the only contexts where collection of sensitive attributes is likely to be justified or important, but they are quintessential areas where such data are already being used to measure and mitigate discrimination. They also highlight major divergences in policy, motivation, and practice.

Comparing these sectors, a complex and inconsistent story emerges. In credit, the law requires some lenders to collect sensitive attribute data, while largely prohibiting others from doing so. In employment, the collection of sensitive attribute data is a familiar part of large employers' day-to-day practice. And in health care, companies' motivation for collecting sensitive attribute data is not just basic antidiscrimination compliance, but rather a moral imperative to address staggering disparities in health outcomes.

We observe that these norms and practices, divergent as they are, typically extend only to traditionally regulated actors. Technology companies that mediate access to opportunities as platforms (e.g., social networks, job boards, and rental sites) or act as vendors to other companies rarely receive clear guidance about when to collect or infer sensitive attribute data. As a result, today, many major technology companies do not collect or infer certain kinds of sensitive attribute data and may therefore struggle to define, detect, and address harms to those protected groups.

We conclude that there are few clear, generally accepted principles about when and why companies should collect sensitive attribute data for antidiscrimination purposes. We emphasize the importance of the machine learning research community engaging on the future development of policy in this area, and urge conversations among stakeholders about whether and how to adapt existing practices or establish new ones.

\subsection{Defining "sensitive attribute data"}

Throughout this paper, we use the term "sensitive attribute data" to refer to details about people's membership in "protected classes" as defined throughout U.S. civil rights laws. This approach to classification is not without its problems: Rigid categories such as these do not currently accommodate nonbinary identities or membership across multiple groups \cite{Hutchinson2019, Hoffman2018, Moss2019, Abdurahman2019}. We acknowledge the reductive and potentially harmful nature of these classification regimes, while simultaneously emphasizing the importance of understanding how they have motivated data collection practices for bias mitigation, and how the history of these practices can inform contemporary contexts.

\subsection{Related work}

The fair ML research community has long reflected on the social and policy contexts of its work, recognizing legal tensions \cite{Barocas2016}, historical parallels in prior debates over definitions of fairness \cite{Hutchinson2019}, and the limitations of data-dependent problem formulation \cite{Selbst2019, Passi2019}. However, when emphasizing the importance of awareness of sensitive attributes in developing and implementing fairness-enhancing interventions, fair ML research and toolkits \cite{Google2018, Pymetrics2019, IBM2019} often take for granted when framing problems and their solutions that sensitive attribute data are available as inputs \cite{Madras2019, Menon2018, Corbett17, Kearns2018, Hardt2016, Kamiran2011, Pedreshi2008}.

When labeled data are not available, researchers have made powerful discoveries by augmenting existing data through the inference or construction of labeled data de novo \cite{Ali2019, Foulds2018, Larson2016, Buolamwini2018}. At times, this work has insufficiently acknowledged the full range of challenges to generating or obtaining those data in applied contexts \cite{Holstein2019, Veale2018, Veale2017}. 

Veale and Binns \cite{Veale2017} and Kilbertus et al \cite{Kilbertus2018} propose approaches to dealing with such information deficits without collecting or revealing sensitive data. Chen et al propose a method to impute unavailable protected class data \cite{Chen2019}. Veale et al \cite{Veale2018} and Holstein et al \cite{Holstein2019} outline the contextual needs for implementing fairness. Zliobaite and Custers use theoretical and linear regression-based examples to argue that sensitive data must be included in the modeling process in order to avoid discrimination \cite{Zliobaite18}. Focusing on the European regulatory environment, their study distinguishes between direct and indirect discrimination, but does not address how different sectoral laws enable or prevent detection of either type.

This paper aims to bridge gaps between theoretical approaches and practical constraints, extracting lessons for fair ML practitioners from three real-world case studies.

\section{Case studies}

\subsection{Credit}

United States federal law prohibits creditors from discriminating on the basis of certain protected characteristics. However, across the credit sector, there are sharply divergent approaches to collecting sensitive attribute data. On one hand, mortgage lenders are required to collect such data from their borrowers. On the other, consumer lenders are largely prohibited from doing so. The reason for this  difference is not immediately apparent, and likely turns on historical details underlying the development of overlapping legal doctrines.

\subsubsection{Background}

In the mid-1970s, policymakers acknowledged that discriminatory practices in the consumer credit and home mortgage industries shut out women, people of color, low-income groups, and others from accessing these vital economic resources. The Equal Credit Opportunity Act (ECOA), passed in 1974, initially made discrimination on the bases of marital status and sex illegal, and was later expanded to include other protected groups. The following year, the Home Mortgage Disclosure Act (HMDA) similarly made discriminating against low-income home mortgage borrowers illegal. Like ECOA, HMDA grew to encompass categories including race, gender, and national origin through subsequent amendments.
 
The origins of ECOA trace back to an era when lenders required unmarried women to have male cosigners on their loans. From the outset, regulators feared that mandatory collection of protected class data beyond gender for the purpose of detecting discriminatory lending might itself facilitate such practices. Thus, under ECOA, the collection of these data is banned. Regulation B, which implements ECOA, has made exceptions for some voluntary collection of data on applicants' color, national origin, religion, race, and sex as "monitoring information" in instances where lenders conduct self-testing to determine whether loans are not being granted to individuals on discriminatory grounds.
 
The Federal Reserve Board (FRB) twice considered amendments to ECOA that would allow voluntary collection of protected class data for non-mortgage loan applicants in order to surface discriminatory lending decisions. In 1995, the first proposal to lift the ban on collecting sensitive attribute data garnered a mix of support and opposition. Noting that discrimination on protected class bases only covered a limited set of criteria for potential disparate treatment during in-person lending scenarios, supporters of the change pointed to the successful identification and reduction of biased mortgage lending decisions that resulted from HMDA's strict data collection practices \cite{Taylor2011}. These advocates disagreed with the FRB's long-held claim that recording these data would lead to discrimination in consumer lending, noting that this predicted harm did not unfold in the home mortgage industry. 

The argument that voluntary collection of protected characteristic information would lead to discriminatory lending persisted, however, in large part due to credit industry representatives' complaint letters opposing the amendments. Banks and other lending institutions were not inclined to support a measure that would incur higher costs and stricter reporting standards and presumably may have revealed discriminatory practices. They additionally warned that being asked about sensitive attributes could deter some minority applicants. In response to these public comments following the proposed amendment in 1995, the FRB decided to leave the decision about collecting protected class data up to Congress.

After introducing a second proposal to remove the ban on collecting these data in 1998, the FRB once again determined in 2003 that consumer lending institutions should not gather this information. Standing by their original conviction that sensitive attribute information collection would lead to outright discrimination, the FRB also reasoned that making this a voluntary action could result in incomplete data collection and inconsistent data formatting that would hinder cross-market comparison between creditors \cite{Taylor2011}.

The ECOA's evolution was in many ways the opposite of HMDA's expansive push to seek evidence of unfair practices in the mortgage lending industry. HMDA grew out of home mortgage depository institutions' disproportionate withdrawal of investments in largely urban areas from which they drew their deposits: a form of redlining that devitalized older neighborhoods, since residents could not access the credit required to sell and refurbish their homes \cite{Kolar2006}. 

HMDA's initial reporting requirements involved publicizing geographic data about lending patterns. As the contexts and causes of home mortgage lending discrimination changed, HMDA was amended between 1980 and the early '00s to expand the scope of institutions covered and to call for reporting of sensitive attribute data on borrowers' gender, race, income, and other categories. When regulators determined these data were insufficient to demonstrate discrimination, they called for further data collection including data about rejected applications and loan pricing.

\subsubsection{Data practices}

While ECOA prohibits collection of sensitive attribute data for most purposes, its implementing regulations allow banks and "anyone who, in the ordinary course of business, regularly participates in decisions about whether or not to extend credit or how much credit to extend" to collect, in a narrow set of circumstances, sensitive attribute data on individuals applying for non-mortgage loans  \cite{Severson2003}. If lenders opt to collect this data, they must indicate that the information is being recorded for self-testing and monitoring purposes. If an applicant prefers not to provide their race and sex information, the lender is allowed to make their own determinations of these characteristics from visual observation and surname analysis. If the self-test demonstrates that the institution may have violated ECOA, the lender must attempt to identify the cause and extent of the violation. Save for in some instances, the results of the self-test are considered privileged information that government agencies cannot access in investigations related to ECOA transgressions.

HMDA, by contrast, requires expansive collection of both sensitive attribute data and related mortgage loan application data that can be used to build arguments that discrimination has occurred. Under HMDA, protected data that must be collected as part of a Loan/Application Register (LAR) include sex, race, and ethnicity, with additional requirements that data on income, loan amount and type, property location, and reasons for loan denial (among others) must be reported \cite{Kolar2006}. Lenders are allowed to use visual observation and surname analysis to guess the sex, race, and ethnicity of applicants who choose not to self-identify these traits. The data are published in different formats depending on the intended recipient. Lenders submit these data to the FRB annually, whereas if a member of the public requested access they would be presented with a modified LAR scrubbed of any identifying information. Finally, the Federal Financial Institutions Examination Council (FFIEC) creates disclosure statements for each lender based on their LAR data, and publishes openly available aggregate reports of HMDA data at city, national, and census-tract levels.

\subsubsection{Results and reactions}

The question of whether sensitive attribute data should be collected to detect discrimination in consumer lending remains controversial. As one scholar put it, "Even if computerized credit scoring arguably has the potential to eliminate disparate treatment results, disparate impact discrimination may still occur" \cite{Taylor2011}. Another scholar has suggested creditors should be {\em required} to conduct self-testing using sensitive attribute data.\cite{Abuhamad2019} Lenders and other proponents of credit scoring systems may argue that expanded collection of data on race and other protected class characteristics would be insufficient to prove discrimination given the increasing complexity of how credit scores are calculated.

Today, as was the case when ECOA was passed, the absence of sensitive attribute data makes it difficult to document and mitigate inequitable consumer lending practices. For example, one of the few, robust public studies on credit scores and discrimination in the United States was performed by the FRB in 2007, at the direction of Congress \cite{FRB2007}. To conduct its analysis, the FRB created a database that, for the first time, combined sensitive attribute data collected by the Social Security Administration (SSA) with a large, nationally representative sample of individuals' credit records. The FRB noted its study was unique in part because of the lack of sensitive attribute data in this domain, and this unusual undertaking would not have been possible without significant governmental time and resources.

The shortage of sensitive attribute data in the consumer lending space also complicates regulatory enforcement. For example, in 2013, the Consumer Financial Protection Bureau (CFPB) and the Department of Justice found that Ally Financial, an auto lending firm, overcharged over 230,000 minority borrowers on their car loans. Two years later, the CFPB required Ally Financial to send checks from its \$80 million settlement to customers believed to have unfairly paid higher prices for their loans \cite{Purtill2015}. Lacking access to data on which exact individuals had overpaid, however, the CFPB instead used a Bayesian Improved Surname Geocoding (BISG) method to predict which customers were likely to be racial minorities, and were therefore more likely to be victims of Ally Financial's allegedly discriminatory pricing. Although BISG's probabilistic means of using publicly available surnames and geographical information as proxies for race and ethnicity is regarded as among the most advanced technique of its kind \cite{CFPB2014}, it is not without flaws. In the use of BISG during the Ally Financial payout, some white Americans were misidentified as having been overcharged for car loans on a discriminatory basis and received compensatory checks \cite{WSJ2015}. Had data collection practices in non-mortgage lending included sensitive attributes, such mistakes could have been averted. Moreover, predictive power of these techniques might diminish over time if housing and marital segregation patterns change.  

By contrast, the amendments to HMDA that spurred collection of protected class data came into effect in 1990, and data from 1992 reflected a significant rise in mortgage lending to low- and moderate-income and minority communities \cite{Marsico1999}. Moreover, in the longer term, the publication of the 1991 data fueled community activism and helped change home mortgage lenders' practices. Making HMDA data mutually accessible to lending institutions and community organizations is correlated with beneficial outcomes for banks and borrowers alike \cite{NCLC2015}. However, it remains difficult to know for certain to what extent this data led to reductions in discriminatory lending practices or merely documented changes that were already underway.

\subsection{Employment}

United States federal law prohibits employers and employment agencies from discriminating on the basis of certain protected attributes. In this context, the collection of demographic information is a familiar part of most employers' day-to-day practice. For example, many large employers are {\em required} to collect demographic data about job applicants and employees to facilitate regulatory enforcement and research. And for many decades, employment selection procedures have been subject to regulatory guidelines that assume "adverse impact" can be readily quantified.

\subsubsection{Background}

Following sustained, nationwide demands to end racial discrimination and segregation, Congress passed sweeping protections in the Civil Rights Act in 1964. Title VII of the Act pertains specifically to employment, prohibiting employers from directly or indirectly discriminating in their employment practices and laying out expectations around data collection and reporting for enforcement purposes. The following year, President Lyndon B. Johnson signed Executive Order 11246, which prohibits federal contractors from discriminating in employment decisions, and also requires employers to take affirmative action to increase the representation of women and minorities in their workforces. The order, enforced by the Department of Labor's Office of Federal Contract Compliance (OFCCP), also outlines related requirements around the documentation of recruitment activities, including the collection of demographic information about job applicants and employees in order to facilitate the detection of discrimination at different points in the recruitment pipeline.

Title VII requires employers and other covered entities to "make and keep such records relevant to the determinations of whether unlawful employment practices have been or are being committed," as defined by the Equal Opportunity Employment Commission (EEOC), which enforces the law \cite{TitleVII}. Since employers may be liable for employment practices that result in disparate impact on the basis of protected categories including race and gender, EEOC guidance points to Title VII as a legal basis for requiring the collection of applicant data as necessary to detect, mitigate, or defend against claims of disparate impact. The Uniform Guidelines on Employment Selection Procedures, which reflects the U.S. government's unified position on employment tests, detail how employment tests must be evaluated for unjustified adverse impact on the basis of race, sex, or ethnicity. The EEOC may allow employers to use selection procedures with disparate impact provided that the procedure has been "validated" according to these guidelines \cite{UniformGuidelines}.

In order to support enforcement of these legal protections, monitor progress in workplace diversity, and enable employer self-assessment, the EEOC also requires private employers with 100 or more employees and contractors with more than 50 employees to collect aggregate statistics about the demographics of their workforce and report them to regulators on a yearly basis, known as EEO-1 reports.

\subsubsection{Data practices}

Collection of sensitive attribute data in the employment sector is highly standardized, reflecting well-defined federal reporting requirements.

For EEO-1 reports, employers must collect data on sex, a binary field (male or female), as well as race, divided into predefined categories of Black, Hispanic, Asian/Pacific Islander, American Indian/Alaskan Native, white, or "two or more races" \cite{EEO2018}. These categories were last updated in 2005 (after 40 years), and in 2007 the EEOC advised that employers were permitted---but not required---to collect more detailed demographic data \cite{Norris2007, OFCCP2008}. Employers must offer employees the opportunity to voluntarily self-identify in the predefined categories. If and only if an employee declines to self-identify, the employer may use "employment records or observer identification," elsewhere described as "visual surveys of the workforce" to categorize the worker to complete their reporting requirements \cite{EEO2018}.

Although not all employers are required to track sensitive attributes from job applicants, many opt to solicit this information at the time of application, and federal contractors are required to do so. Contractors may solicit demographic data from applicants at any time during the employee selection process so long as the data is solicited from all applicants. Regulators advise that "voluntary self-reporting or self-identification is still generally the preferred method for collecting data on race, ethnicity, and gender, but in situations where self-reporting is not practicable or feasible, observer information may be used to identify race, ethnicity, and gender" \cite{InternetApplicant}. After making "reasonable efforts to identify applicant gender, race, and ethnicity information," contractors may record the applicant's race and gender as "unknown"---with the exception that employers may visually identify applicants "when the applicant appears in person and declines to self-identify" \cite{OFCCP2005}. Notably, employers may not use these data as a part of their employment selection procedures, but may use them to evaluate outcomes and inform changes to those procedures.  

\subsubsection{Results and reactions}

As of 2017, nearly 70 thousand employers file EEO-1 reports per year, documenting data for over 50 million employees \cite{JobPatterns}. Multiple studies have used EEO and other sources of demographic data to measure trends in occupational segregation, finding that it has declined since the passage of Title VII \cite{Robinson2005, Weeden2018, Kurtulus2012}. Others use this data to more closely examine race and sex inequality in managerial positions and within specific industries, as well as gender and racial pay gaps \cite{Robinson2005, Tomaskovic2012, Huffman2010}. Several researchers were able to determine that OFCCP monitoring and enforcement in particular likely contributed to greater representation of Black workers in skilled occupations \cite{Leonard1984}. The EEOC and OFCCP themselves commonly use EEO-1 and other mandatorily collected data to support investigations of individual and systemic employment discrimination \cite{EEOC2016, OFFCPNumbers}.

Some have pointed out that unlike other government survey instruments, the EEOC merges data on race and ethnicity, which may lead to measurement errors \cite{Robinson2005}. Others critique the allowance of observed data, but concede that because observed data relate to how workers may be perceived, these data may still have utility in understanding employment discrimination \cite{Simon2005}. However, we identified relatively little criticsm of the overall exercise of collecting sensitive attribute data in the context of employment, perhaps because the law requiring and justifying their collection is so clear.

Here again, it is not clear that the relationship between demographic data collection and any occupational desegregation is a causal one. Without this disaggregated employment data, however, documentation of these trends would be significantly more difficult. Indeed, researchers have found that while EEO-1 data do have some constraints, they can be a particularly powerful tool to study workplace inequality and segregation, especially as compared with other data sources \cite{Robinson2005}.

\subsection{Health}

United States federal law prohibits discrimination in the provision of various health care services. For example, those who qualify for federal health insurance programs such as Medicare or Medicaid may not be subjected to discrimination based on certain sensitive attributes. However, unlike in credit and employment, a major driving factor behind collection of sensitive attribute data in this sector has been voluntary industry efforts to address racial and ethnic disparities in health outcomes, rather than compliance with antidiscrimination laws alone.

\subsubsection{Background}

The passage of the Civil Rights Act in 1964 and the establishment of Medicare the following year created a need for data to confirm that patients had equal access to health care and that hospitals were not segregated. As a result, many hospitals initially collected data about sensitive attributes for compliance purposes only \cite{AHRQCategorization}.

A shift in approach was prompted not long after by Secretary of Health and Human Services (HHS) Margaret Heckler's observation in a 1983 national health report that minority health lagged behind that of white Americans, and the subsequent formation of the Task Force on Black and Minority Health to research this gap. The 1986 publication of the Report of the Secretary's Task Force on Black and Minority Health (Heckler Report) marked the first study highlighting the significant health disparities racial minorities experienced in the U.S. \cite{HecklerReport}. Although the Heckler Report's findings drew awareness to racial inequality in healthcare provision, they did not themselves effect a shift away from compliance-based sensitive attribute data collection toward a model of using these data to reduce discrimination. 

At the request of Congress in 2003, the Institute of Medicine (IOM) published a follow-up report, Unequal Treatment, affirming that unacceptable levels of racial and ethnic disparities in health outcomes persisted \cite{IOMReport}. The IOM report concluded that without data on patients' race, ethnicity, socioeconomic status, and primary language, it would be impossible for healthcare providers to detect or address these disparities, and recommended the systematic collection and reporting of race and ethnicity data as a critical step toward eliminating them.

The IOM report jump-started health insurance and other care providers' joint, voluntary effort to collect and use data for healthcare quality improvement and disparity reduction \cite{AHRQCategorization}. Organizations such as the National Health Plan Collaborative (NHPC) connected health research institutes to national and regional health plans in order for the former to provide these firms with educational tools and recommendations for how to detect and mitigate discrimination \cite{NHPC2009}. While initially, many insurance providers believed collecting race and ethnicity data was illegal, legal analysis determined that collection was justified under (though not explicitly required by) Title VI of the Civil Rights Act and the Affordable Care Act, as well as several state laws. Under these statutes, health plans are prohibited from using demographic data for discriminatory purposes, including steering patients toward certain healthcare products \cite{Kornblet2008, NRC2004}. However, health plans are allowed to use these data in order to report aggregate trends and join initiatives to provide equitable services.

\subsubsection{Data practices}

Some health providers have found it necessary to collect data on patients' race, ethnicity, and primary spoken language (REL) to identify health care disparities \cite{AHRQ2012}. However, there is substantial variability in the precise categories and level of granularity different health providers opt to use to do so. Industry-wide efforts to standardize these data are ongoing.

Physicians and hospitals often collect REL data at intake---usually by asking patients directly, though sometimes determined by intake specialist observation \cite{RWJF2011}. Health plans, on the other hand, tend to use surveys and incentive programs to collect data after people have signed up for coverage. In some cases, insurers are prohibited from asking for race/ethnicity data during the sign-up process \cite{RAND2016, AHRQSummary}. Some health providers also appear to be able to share and obtain data from federal agencies (e.g. Medicaid), though the exact mechanics of this process remain obscure.

Policymakers and practitioners recognize that in general, data that patients self-report are strongly preferred \cite{Hasnain2006, AHRQIntro}, but in practice, providers have struggled to convince most patients to voluntarily self-report. In the interest of generating data necessary to reduce disparities, methods to estimate race and ethnicity have been widely adopted to supplement self-reported data \cite{Nerenz2013}. Early inference methods involved basic geocoding and surname analysis; more advanced probabilistic techniques have since been developed to refine these estimates. These algorithms produce probabilities that individuals belong to a particular racial or ethnic group, which can then be used to assess disparities between subgroups at an aggregate level \cite{RAND2016, AHRQCategorization}. A number of health plans combine self-reported and estimated data to increase accuracy of their analysis \cite{Nerenz2013}.

Experts have recommended that race/ethnicity data based on indirect estimation methods should be stored separately from or be clearly marked in medical systems. Inferred data should not be placed in individuals' clinical medical records---that is, probabilistic methods should not be used to assign someone a particular race or ethnicity classification \cite{AHRQCategorization}---but should only be used for aggregate statistical analysis \cite{AHRQSummary}. The IOM recommended that when possible, estimations should be accompanied by their respective probabilities \cite{AHRQCategorization}. Whether actual data management practice follows these recommendations likely varies by institution.

\subsubsection{Results and reactions}

While significant healthcare disparities remain, they have narrowed since the publication of the IOM report that motivated increased data collection \cite{IOM2012, NCHS2016}. Moreover, granular data has enabled ongoing monitoring and benchmarking of health outcomes, motivated substantial scientific and policy research, and supported federal, state, local, industry, and practitioner-driven disparity reduction initiatives. 

Although many health plans have internal policies on confidentiality and use of race/ethnicity data \cite{Gazmararian2012}, low rates of participation in voluntary data collection may indicate continued lack of trust in healthcare institutions that collect these data, and fear that demographic data might be used to discriminate against patients or otherwise be misused \cite{Hasnain2006}. Health plans have admitted that they sometimes hesitate to collect data for fear of being accused of discrimination \cite{AHRQSummary}, on top of other challenges like privacy concerns, IT limitations, and inconsistency or insensitivity in the available categories \cite{Nerenz2013}. But many healthcare providers circumvent these challenges by using techniques to generate demographic data in a probabilistic manner. 

Critiques of direct and indirect data collection efforts in healthcare have also emerged on the grounds that concepts of race and ethnicity are merely sociopolitical constructs \cite{AHRQSummary}, and therefore categorizing patients using those constructs may reinforce and calcify them. However, the broadly recognized harms of race- and ethnicity-related health disparities seem to have outweighed this critical perspective for the time being.

\section{Discussion}

Clearly, debates about collection of sensitive attribute data for antidiscrimination purposes are not new. There are decades of precedent that can inform the machine learning fairness research community, the broader technology industry, and other stakeholders.

It is important to reiterate that our case studies do not indicate whether collection of sensitive attribute data has contributed causally to more fair and equitable outcomes. A more fulsome analysis of this question remains for future work. However, we remain convinced that measurement is often a precondition for meaningful improvements.

While the case studies above merely scratch the surface, they offer some important insights. First, they show that U.S. legal frameworks do not offer consistent, extensible guidance about when and how corporations should collect sensitive attribute data. Rather, there are divergent and sometimes contradictory approaches: Some companies are required to collect sensitive data to comply with antidiscrimination laws, while others are explicitly prohibited from doing so. Second, they show that companies' primary incentives for collecting sensitive attribute data may not---and need not---be compliance or legal requirements at all. The healthcare industry is one such example. Here, deliberate, sustained, and ongoing debates on data collection and inference practices across the industry and stakeholder communities were needed to align on an approach to combating disparities.

If awareness-based techniques remain a primary approach to bias mitigation in predictive modelling, there is a need to thoughtfully consider what efforts must be undertaken to expand collection of sensitive attribute data in a responsible manner.

\subsection{Lessons regarding traditionally regulated companies}

For traditionally regulated entities like banks and employers, modernization or clarification of laws and regulatory guidance may be needed to encourage the collection of sensitive attribute data for new antidiscrimination efforts. Because these companies can be liable for discriminatory outcomes, they are unlikely to voluntarily collect or analyze sensitive attribute data that could introduce new vectors for liability. Thus, they might resist legal reforms that make it easier to collect sensitive attribute data.

Looking ahead, policymakers, researchers, and and civil society will need to work together to assess what kinds of sensitive attribute data are needed to protect people against discrimination {\em and} create the policy conditions for that collection to occur. These stakeholders will need to consider what data ought to be collected and in what form, and the appropriate scope of "safe harbor" provisions to incentivize thorough and transparent study. These are not clear or settled questions, even with decades of practice under longstanding civil rights laws.

\subsection{Lessons regarding less regulated companies}

Many technology platform companies, including those using models to mediate access to important life opportunities, are not squarely covered by civil rights laws. These companies often operate as internet intermediaries, and thus enjoy some special legal protections from liability arising from content posted by third party users \cite{Section230}. Nonetheless, many are grappling with how to prevent bias. For example:

\begin{itemize}
\item Airbnb recently assembled "a permanent team of engineers, data scientists, researchers, and designers whose sole purpose is to advance belonging and inclusion and to root out bias" \cite{Airbnb2016}. The announcement came on the heels of reports of discrimination against African Americans seeking housing opportunities on its platform. The company has not yet publicly discussed the details of this work, or whether it collects or infers sensitive attribute data in its efforts to combat discrimination. However, it is difficult to imagine an approach that would avoid these questions.
\item Facebook, in delivering advertisements on its platform, introduces demographic skews along gender and race lines \cite{Ali2019}. This practice is currently being challenged in court by the United States Department of Housing and Urban Development (HUD) \cite{HUD2018}. Furthermore, as part of a legally enforceable settlement with civil rights organizations, Facebook recently committed to studying the potential for unintended biases in algorithmic modeling \cite{NFHA2019}. However, this research will likely be impossible without collecting or inferring sensitive attributes of the company's users. It is not yet clear how Facebook will approach this issue.
\item LinkedIn, in an effort to promote equity in hiring, recently updated its recruiter tools to balance the gender distribution in candidate search results, rather than sorting candidates purely by "relevance" \cite{BusinessInsider}. With this update, if the pool of potential candidates who fit an employer's search parameters reflects a certain proportion of women, LinkedIn will re-rank candidates so that every page of search results reflects that proportion. The company also plans to offer employers reports that track the gender breakdown of their candidates across several stages of the recruitment process, as well as comparisons to the gender makeup of peer companies. These features rely on inferring gender data about jobseekers on the platform, which the company was already doing for advertising purposes.
\end{itemize}

It's not surprising that each of the above examples was motivated by some combination of public pressure or litigation. Technology companies are unsure about what kinds of sensitive attribute collection are appropriate. As a result, the path of least resistance is to simply not to collect or infer data that may create controversy or highlight disparities that may be difficult to address. This is especially true given that perceived violations of privacy are likely to garner intensive media coverage, or where applicable, increased attention from regulators. It will likely fall to a wide range of stakeholders, including advocates, researchers, and policymakers, to ensure that sensitive attribute data is collected and used under appropriate circumstances.

\subsection{The need for multidisciplinary collaboration}

The implementation of awareness-based antidiscrimination approaches cannot, and should not, move forward without robust involvement of public interest, technical, and regulatory stakeholders. Even amid clear and compelling risks of discrimination or unjust demographic disparities, it can be difficult for policymakers to recommend the collection of sensitive attribute data. There is no evidence this issue will become easier in the future, despite the rapid adoption of machine learning models involved in important life decisions for which these data may be critical to prevent harm. 

Privacy laws can sometimes sit in tension with antidiscrimination goals, and might prevent well-meaning actors from collecting data that are necessary to detect and remediate bias in machine learning-based models. Privacy advocates will need to ensure that new legal requirements around data minimization and restrictions on the processing of sensitive data do not deter or impede companies from good faith self-testing and bias remediation. At least one recent U.S. legislative proposal provides an explicit exception for such testing \cite{COPRA}, reinforcing the need for more detailed implementation guidance. Meanwhile, European laws and norms diverge significantly from the U.S. approach, prioritizing privacy heavily over awareness-based antidiscrimination approaches \cite{Zliobaite18}. Private entities may need to navigate conflicting laws, guidance, and public expectations across social and geopolitical contexts.

Finally, there is no shortage of critical questions that still need to be answered:

\begin{itemize}
\item \emph{When should sensitive attribute data be collected?} Given the practices described above, non-industry actors should consider under what conditions, if any, they would trust certain private actors with sensitive attribute data that are needed for antidiscrimination efforts. It's obvious that data collection would be justified in some contexts, but the risks may outweigh potential benefits in others. It's far less obvious (and beyond the scope of this paper to suggest) where those lines should be drawn. These norms are especially unsettled for technology companies, who have not had the same historical obligations as traditionally regulated entities, and suffer from significant trust deficits around their data practices.

\item \emph{How should sensitive attribute data be created?} Sensitive attribute data can be collected directly from subjects or inferred from non-sensitive data. However, inference presents challenges around consent, forced classification, and error. Stakeholders must work together to determine under what conditions inference is acceptable, appropriate inference methodologies, and how to treat inferred data responsibly. The cases considered in this paper offer instructive approaches, including retaining probabilistic values and uncertainty in inferred data, clearly marking when data are observed or inferred, and storing inferred data separately from data collected with permission. Other approaches might include enforceable commitments to use these data only for detecting and mitigating discrimination.

\item \emph{How should sensitive attribute data be treated and secured?} Ideally, sensitive data would be stored separately from other data and used only for limited purposes, but such technical safeguards may be difficult to guarantee. New privacy-protective techniques to access sensitive attribute data, including secure multi-party computation tools like private set intersection and homomorphic encryption, may allow companies to securely sequester these sensitive data from general purpose user data, or even enable trusted third parties to collect, infer, or hold sensitive data while making their insights available to the private entities whose products implicate people's rights \cite{Veale2017}. However, these techniques are still nascent and have yet to be broadly deployed for the purpose of bias testing or mitigation.
\end{itemize}

\section{Conclusion}

Policy debates about the collection and use of sensitive attribute data will decide the fate of awareness-based bias mitigation techniques. There is an urgent need for machine learning scholars to drive these conversations forward, along with other stakeholders, so policy and technical approaches can be developed in accordance with each other. The ability to detect and address bias in algorithms---and the durability of foundational civil rights protections---may hang in the balance.

\bibliographystyle{ACM-Reference-Format}
\bibliography{bibliography}

\end{document}